# How the instant collapse of a spatially-extended quantum state is consistent with relativity of simultaneity

Moses Fayngold
Department of Physics, New Jersey Institute of Technology, Newark, NJ 07102

Intrinsically probabilistic nature of quantum phenomena is what makes Quantum Mechanics consistent with relativistic postulates. We analyze a simple thought experiment demonstrating intimate connection between Quantum Mechanics and Special Relativity. The experiment involves collapse of an extended quantum state of a single particle under position measurement. Such collapse is instantaneous for all observers according to Quantum Mechanics but must generally be a lasting process according to Relativity. Apparent clash between these two statements is eliminated by probabilistic nature of the Born postulate. The consistence of Quantum Mechanics with Relativity is also shown for entangled composite systems.

## 1. Introduction

In standard descriptions of quantum measurement (see, e.g., [1]), the initial state instantly "collapses" to an eigenstate of the measured observable. In a position measurement, a spatially extended state $\Psi(\mathbf{r},t)$ shrinks immediately to a point-like eigenstate of position operator $\hat{\mathbf{r}}=\mathbf{r}$: $\Psi(\mathbf{r},\,t)\to\delta(\mathbf{r}-\mathbf{r}')$, and this transition is instantaneous. The conventionally used term "collapse" for such process is misleading since it totally ignores the unavoidable accompanying effect of spreading of the corresponding momentum spectrum. It is more appropriate to call the whole effect the "quantum jump" or "instant reconfiguration" looking like collapse in one representation and like an explosion (micro-analog of the Big Bang) in another one [2]. What is really important here is the word "instant" reflecting the fact that this process takes, in contrast with continuous evolution, no time: it is instantaneous in *any* reference frame (RF). This seems to take us back to pre-relativistic notion of absolute time, when a group of simultaneous events in one RF is simultaneous in all others. But according to special relativity (SR), spatially separated events simultaneous in one RF are generally time-separated in others. In the language of mathematical formalism of SR, simultaneity is not invariant under rotations in space-time. Thus, delocalized quantum states forming the backbone of quantum mechanics (QM) [1-13] may appear to clash with SR.

But the consistence between QM and SR is well-established (see [14] and references therein). Here we demonstrate this consistence by analyzing a simple thought experiment, which requires only rudimentary knowledge of QM such as the superposition principle and the Born rule $\mathcal{P}(\mathbf{r},\mathrm{t})=\left|\Psi(\mathbf{r},t)\right|^2$ for probability density of an event.

In Sec. 2, we consider collapse of an extended state of a single particle under position measurement as observed from two different inertial RF. In Sec. 3, the collapse of an entangled state of a composite system is discussed, also from the viewpoint of different observers. In both cases, the analysis unveils the factors that preclude any conflicts between QM and SR.

## 2. Quantum collapse in position measurement

Consider a *single* particle in a spatially-extended state $\left|\Psi\right\rangle$ in a free space. The position indeterminacy in such states is manifestation of quantum non-locality (QNL), although this term usually refers to entangled *composite* systems. For a single particle, QNL is manifest through continuous superposition of eigenstates $\left|\mathbf{r}'\right\rangle$ with definite coordinates $\mathbf{r}'$:

$$\left|\Psi\right\rangle=\int\chi(\mathbf{r}')\left|\mathbf{r}'\right\rangle d\mathbf{r}' \qquad (2.1)$$

Explicit position representation is given by projection of (2.1) onto a state $\left|\mathbf{r}\right\rangle$:

$$\left\langle\mathbf{r}\middle|\Psi\right\rangle=\Psi(\mathbf{r}),\quad \left\langle\mathbf{r}\middle|\mathbf{r}'\right\rangle=\delta(\mathbf{r}-\mathbf{r}'),\ \chi(\mathbf{r}')=\Psi(\mathbf{r}') \qquad (2.2)$$

This takes us to familiar

$$\Psi(\mathbf{r})=\int\Psi(\mathbf{r}')\,\delta(\mathbf{r}-\mathbf{r}')\,d\mathbf{r}' \qquad (2.3)$$

In an idealized model, the position measurement collapses state (2.1) to some definite location $|\mathbf{r}\rangle$:

$$|\Psi\rangle = \int \chi(\mathbf{r}')|\mathbf{r}'\rangle d\mathbf{r}' \quad \Rightarrow \quad |\mathbf{r}\rangle \qquad (2.4)$$

The corresponding outcome – finding the particle in the vicinity of $|\mathbf{r}\rangle$ – can be predicted only probabilistically, with probability $d\mathcal{P}(\mathbf{r}) = |\Psi(\mathbf{r})|^2 d\mathbf{r}$. And there are *no information transfers* between different parts of packet (2.1) in the process of its collapse.

The widely used term "*collapse*" here and generally in QM is largely misleading. In case of position measurements, it brings to mind some fluid-like entity instantly converging to a point, something closely imitating gravitational collapse. This is a false analogy. Gravitational collapse is the convergence of matter with finite speed [15, 16] and can last arbitrarily long by clocks of a distant observer [17]. Most importantly, it is a deterministic process caused by gravitational attraction to the respective source and is described by equations of General Relativity. In contrast, a QM state collapses instantly, without any convergence of its parts, and does not have any pre-existing center of attraction. It can collapse to any point which cannot be predicted in advance. This process is not described by any equation. (Some features of the QM collapse are treated by the decoherence theory [18-20], but they are irrelevant to the basic point of this article).

One of the most crucial aspects of QM "collapse" is the unavoidably accompanying *opposite* effect that could (also inadequately!) be called "explosion" [2]. For example, "collapse" of an extended state in configuration space is accompanied by its "explosion" in momentum space, and vice versa.

The fundamental dualism "collapse-explosion" in a QM measurement is overlooked in most treatments on QM. It is hard to find a term describing all features of QM "collapse" in one word. We will use *instant reconfiguration* (IR) as a far better term.

Now we turn closer to the physical aspects of IR and analyze them in more details. The goal will be to demonstrate the consistency of QM with SR in the most-simple way.

Consider a thought experiment of position measurement for a single electron from a quasi-monochromatic source. According to (2.1), such an electron can be crudely visualized as "smeared out" over the space. If it had been fired with definite momentum, it will be described by de Broglie's wave. We assume that quantum noise can be neglected.

The position measurement scheme can be a set of detectors in the electron's way. In the simplest case of de Broglie's wave, there will be an equal chance for the electron to be recorded by any detector. When this happens, denote the respective coordinate and time as (0, 0). Now the pre-existing state is reconfigured to definite location – that at the origin of our RF. And once the electron is with certainty there, the respective probability instantly jumps from $d\mathcal{P}(0) << 1$ to $d\mathcal{P}(0) = 1$. Accordingly, since the electron now cannot be anywhere else, all probabilities outside the detector vanish at once, $\mathcal{P}(\mathbf{r}) = 0$ for all $\mathbf{r} \neq 0$. The whole state undergoes IR

$$\Psi(\mathbf{r}) \;\Rightarrow\; \delta(\mathbf{r}), \qquad (2.5)$$

which is just expression (2.4) in position representation. The net probability is conserved, but the state has dramatically changed. And this will hold in any RF.

Denote the original RF as K and consider another frame K' moving relative to K with velocity **V**. Residents of K' also observe the IR of the wave packet. Suppose, at the moment when the origins of K and K' coincide, their respective *local* clocks both read $t = t' = 0$. Then the initial packet disappears in $\mathrm{K}'$ instantly at $t' = 0$ everywhere except for the origin, where the probability jumps to 1. The picture is identical to that in K.

But according to SR, the events spatially separated along **V** and simultaneous in K are not simultaneous in K', and vice versa. Consider a huge air-bubble in vacuum extended along **V** and instantly losing all its air when pricked in many places at once in K. Imagine also additional small bubble inside, immediately pumped up with the equal amount of air. This might look as a crude model of IR with conservation of the net probability. But, whereas IR is instantaneous in *all* RF, the presented process as observed in K' is a time-extended succession of pricks and respective local air-losses. The remaining feature common with K is that pricks are not causally connected [21, 22].

In contrast with this failing analogy, the electron's initial "omnipresence", albeit a real physical characteristic, is spread probability rather than matter. This crucial difference allows IR to be an instant process in all RF and yet remain consistent with SR.

To see intimate connection between SR and probabilistic nature of QM, consider a possible experimental scheme of position measurement. Imagine a row of detectors in K along direction of **V**. The particle can be found in one of detectors while all others will remain idle. Denote position and time coordinates of the clicked detector as (0, 0).

Now consider this process from K'. *The same* common moment of *remaining idle* in K at $t = 0$ corresponds to *different* moments in K'. But since nothing really happens with all idle detectors, their *inactions* simultaneous in K can be considered as simultaneous in K' as well. The deep physical reason for it is that practically any change of local probability is not a directly observable event. The only exception is jump of one of probabilities to 1, which immediately manifests itself in firing of the respective detector. Recording all other detectors simultaneously with the one that clicked makes no physical difference from their recording at successive moments of local times. It is perfectly safe to say that the set of local *probabilities* (and thereby the physical state described by them!) changes instantly for all observers. Residents of all RF will record only one event – the click of detector $D_0$ at (0, 0). The fact that this detector is moving in all RF except for K, is totally immaterial for the presented argument. We could as well consider a set of detectors stationary in $\mathrm{K}'$ - with the same result.

It has been argued that the absence of a click in a detector, being a possible outcome of an experiment, "*…is also an event and is part of the historical record*" [14].

Recording the *absence* of something does not bring this something in. An *event* is an *observable rapid change* at a given location. The absence of a click in a given detector is not informative even if combined with the signal from the *clicked* detector $D_0$. Such signal will come only *after* the time $t(\mathbf{r}_i) = |\mathbf{r}_i| / c$, where $\mathbf{r}_i$ is separation between the two detectors. When such signal comes, the respective local observer knows without looking that there are no clicks in his/her detector or elsewhere except for (0, 0) (noise neglected).

Technically, the set of all considered "non-events" numbered in K simultaneously with event (0, 0) is not simultaneous in $\mathrm{K}'$, and vice-versa. But due to being eventless except for one common element (0, 0), both sets are physically identical, so *each RF is equally entitled to identify its own set at one moment of that frame*. In other words, IR happens on the corresponding *simultaneity section* (*cut in space-time containing event* (0, 0)) for each RF. Geometrically, the cuts are different. But for *probabilities*, the world is insensitive to choice of the cut in the discussed process. This makes QM collapse an instant process in *any* RF without contradicting SR (see also [23]).

Thus, the collapse is instantaneous in different RF because the respective experimental setups (systems of synchronized clocks etc.) automatically select the respective *simultaneous cuts* through space-time. The corresponding records are compatible only because the collapsed entity had been a *probability* cloud, and because it has collapsed to a single location. A hypothetical collapse of a single particle state to two separate locations at once would mean a different world with no conservation laws.

This analysis shows that QM which embraces QNL as its crucial feature can be in vibrant and fruitful harmony with SR only by being intrinsically probabilistic.

With all that, we must be aware of the borderline in the above argument. It works only for IR, not for *continuous evolution* of the state. In the latter case, the probability amplitude written in K as $\Psi(x,t)$ is a different function of coordinates in $K'$. For a scalar function $\Psi(x,t)$ the corresponding Lorentz transformation gives

$$\Psi(x,t) \Rightarrow \tilde{\Psi}(x',t') = \Psi\left( \gamma(V)\left(x'+Vt'\right),\ \gamma(V)\left(t'+\frac{V}{c^2}x'\right)\right), \qquad (2.6)$$

where $\gamma(V)$ is the Lorentz-factor. The state looks differently in different RF. For example, it may be a wave packet with different group velocities $\mathbf{v}(\mathbf{k}) \neq \mathbf{v}'(\mathbf{k}')$ in K and $K'$, respectively. And this is an important, physically observable difference even though it is still only difference of probabilities. The reason is rather subtle. Continuous change of local probability under wave equation still does not have a full status of an event. But in contrast with a single measurement on one particle, it can be observed indirectly in a set of such measurements on a pure ensemble of particles.

In a single position measurement, we only find the particle at some location. The result does not reveal the pre-existing probability distribution. In the same measurement on a pure ensemble, we find the particle's probability distribution as a function of $\mathbf{r}$ and $t$. This is an important characteristic of the state, and it may have different "faces" in different RF, as is evident from (2.6).

A particle described by a wave packet is manifestation of *position indeterminacy*. Once such indeterminacy is accepted as a basic feature of the world, QNL is there already for a single particle. And the corresponding state can evolve either continuously under wave equation, or discontinuously under measurement, when the whole packet instantly disappears and reemerges at the same moment as a narrow spike.

A subtle connection of $\mathcal{P}(\mathbf{r})$ with reality is seen, e.g., in the product $\rho_e(\mathbf{r}) \equiv q_e \mathcal{P}(\mathbf{r})$, where $q_e$ is the electron charge. It determines the effective charge density and hence the electric current in an atom and the corresponding magnetic field produced, say, by an atomic electron with orbital angular momentum $L$. The distribution $\rho_e(\mathbf{r})$ can be treated as a charged "cloud" in the corresponding calculations [2, 23, 24], but undergoes IR together with $\mathcal{P}(\mathbf{r})$ in a position measurement. One might therefore argue that in contrast with $\mathcal{P}(\mathbf{r})$, the instant disappearance of the $\rho_e(\mathbf{r})$ at a given location must be an *observable event* that could be immediately recorded by the respective detector. But that would be a wrong argument. The electron charge (let alone its fraction!) is observable only through its electromagnetic (EM) field. Any change of the field in vacuum propagates at the speed of light, so even at the IR, the corresponding field reshapes only gradually; and all that – in the classical limit of QM for a system with a charge $Q \gg q_e$. The field of a single electron is represented by virtual photons, whose fleeting appearances are never sufficient for their actual detection [25- 27]. So $\rho_e(\mathbf{r})$, while being an objective characteristics of a state, can be only indirectly observed and thus also satisfies the conditions of the considered thought experiment. Therefore such experiment demonstrates general consistency between QM and SR for 1-partite states.

### 3. Instant reconfiguration of entangled states under measurement

Here we will show the consistency between QM and SR for a *composite* system. We consider the simplest case – a pair of separated entangled particles A and B, but the presented argument will hold in general. Let A have two eigenstates $|\sigma_i\rangle_{\mathrm{A}}$ with eigenvalues $\sigma_i$ ($i=1,\ 2$) of observable $\sigma$, and similar for B. System (AB) resides in a 4-dimensional (4D) Hilbert space $\mathcal{H}^{(4)}$. Its most appropriate ortho-normal basis is formed by 4 product states $|\sigma_i\rangle_{\mathrm{A}}|\sigma_j\rangle_{\mathrm{B}}$. An *entangled* state occupies a 2D subspace $\mathcal{H}^{(2)}$, e.g.,

$$|\Psi\rangle_{\mathrm{A,B}} = c_1|\sigma_1\rangle_{\mathrm{A}}|\sigma_1\rangle_{\mathrm{B}} + c_2|\sigma_2\rangle_{\mathrm{A}}|\sigma_2\rangle_{\mathrm{B}} \qquad (3.1)$$

or

$$|\Psi\rangle_{\mathrm{A,B}} = c_1|\sigma_1\rangle_{\mathrm{A}}|\sigma_2\rangle_{\mathrm{B}} + c_2|\sigma_2\rangle_{\mathrm{A}}|\sigma_1\rangle_{\mathrm{B}} \qquad (3.2)$$

with $|c_1|^2 + |c_2|^2 = 1$. The subscript $\alpha = 1,\ 2$ labeling amplitudes $c_\alpha$ is independent from *i*, *j*, labeling eigenstates. State (3.1) can describe, e.g., two circularly polarized photons flying apart in the opposite directions, with the zero net spin. In circular polarization basis $|\sigma_{1,\,2}\rangle = |L\rangle, |R\rangle$ (left- and right-polarized state, respectively), (3.1) will read

$$|\Psi\rangle_{\mathrm{A,B}} = c_1|L\rangle_{\mathrm{A}}|L\rangle_{\mathrm{B}} + c_2|R\rangle_{\mathrm{A}}|R\rangle_{\mathrm{B}} \qquad (3.3)$$

Actual spins in each term here are opposite to each other to make the net spin zero. They are formally equal only with respect to the corresponding momentum vectors which are opposite to each other.

The case (3.2) may be a pair of entangled electrons with the net spin $\mathbf{S}=0$. In the basis $|\sigma_{1,\,2}\rangle = |\uparrow\rangle,\ |\downarrow\rangle$ ("spin-up" and "spin-down", respectively), (3.2) reads

$$|\Psi\rangle_{\mathrm{A,B}} = c_1|\uparrow\rangle_{\mathrm{A}}|\downarrow\rangle_{\mathrm{B}} + c_2|\downarrow\rangle_{\mathrm{A}}|\uparrow\rangle_{\mathrm{B}} \qquad (3.4)$$

The considered entangled systems have the following properties:

(I) Superposed states of (AB) are strictly correlated (cases (3.1), (3.3)), with both – A and B – showing the same individual measurement outcomes, or anti-correlated (cases (3.2), (3.4)), with A and B collapsing to opposite individual states. It has been suggested to call such states the "*superposition of correlations*" or *anti-correlations*, respectively [28]. We find it more convenient to name these varieties of correlations the (+)correlations for states like (3.1, 3), and $(-)$ correlations for states like (3.2, 4). Mathematically, the most general property of an *entangled* state is *inseparability*: it cannot be represented as the product of individual states.

(II) The entanglement is maximal at $|c_1| = |c_2|$. Introducing the ratio $\eta \equiv |c_1|/|c_2|$, we can write this condition as $\eta = 1$. The farther is the system from this condition, the weaker the entanglement. It disappears when $\eta = 0$ or $\eta \to \infty$, that is, when one of the $c_\alpha$ is zero (disentanglement). The latter happens in a $\sigma$-measurement.

(III) Correlation generally weakens if we switch from basis $|\sigma_i\rangle$ to $|\tau_j\rangle$, where observable $\tau$ is incompatible with $\sigma$, e.g. when $\sigma$ and $\tau$ are different spin components $\sigma = s_z$, $\tau = s_x$. Then

$$|\sigma_1\rangle = \frac{1}{\sqrt{2}}\left(|\tau_1\rangle - i|\tau_2\rangle\right); \quad |\sigma_2\rangle = \frac{1}{\sqrt{2}}\left(|\tau_1\rangle + i|\tau_2\rangle\right) \qquad (3.5)$$

Putting this, say, into (3.1) yields

$$|\Psi\rangle_{\mathrm{A,B}} = \frac{1}{2}\left\{(c_1 + c_2)\left(|\tau_1\rangle_\mathrm{A}|\tau_1\rangle_\mathrm{B} - |\tau_2\rangle_\mathrm{A}|\tau_2\rangle_\mathrm{B}\right) - i(c_1 - c_2)\left(|\tau_1\rangle_\mathrm{A}|\tau_2\rangle_\mathrm{B} + |\tau_2\rangle_\mathrm{A}|\tau_1\rangle_\mathrm{B}\right)\right\} \quad (3.6)$$

Here *any* of the 4 product states $|\tau_i\rangle_\mathrm{A}|\tau_j\rangle_\mathrm{B}$ can generally appear in a $\tau$-measurement, but their probabilities depend on $c_\alpha$ . At maximal entanglement, (3.6) may reduce to

$$|\Psi\rangle_{\mathrm{A,B}} \Rightarrow \frac{1}{\sqrt{2}}\begin{cases} |\tau_1\rangle_\mathrm{A}|\tau_1\rangle_\mathrm{B} - |\tau_2\rangle_\mathrm{A}|\tau_2\rangle_\mathrm{B}, \quad c_1 = c_2 & (3.7) \\ \text{or} & \\ -i\left(|\tau_1\rangle_\mathrm{A}|\tau_2\rangle_\mathrm{B} + |\tau_2\rangle_\mathrm{A}|\tau_1\rangle_\mathrm{B}\right), \quad c_1 = -c_2 & (3.8) \end{cases}$$

In this case the particles' states are strictly $(+)$ or $(-)$ correlated with respect to $\tau$ as well. The opposite case, with one of the amplitudes $c_\alpha = 0$, is a disentangled state:

$$|\Psi\rangle_{\mathrm{A,B}} \Rightarrow \begin{cases} |\sigma_1\rangle_A|\sigma_1\rangle_B = (1/2)\left(|\tau_1\rangle_A - i|\tau_2\rangle_A\right)\left(|\tau_1\rangle_B - i|\tau_2\rangle_B\right), \quad c_2 = 0 & (3.9) \\ \text{or} & \\ |\sigma_2\rangle_A|\sigma_2\rangle_B = (1/2)\left(|\tau_1\rangle_A + i|\tau_2\rangle_A\right)\left(|\tau_1\rangle_B + i|\tau_2\rangle_B\right), \quad c_1 = 0 & (3.10) \end{cases}$$

This state, albeit *disentangled*, is totally correlated in the $|\sigma_i\rangle$-basis but totally uncorrelated in the $|\tau_i\rangle$-basis. And the same may hold for $(-)$ correlations. On the other hand, entangled state (3.6) is totally uncorrelated at

$$|c_1 + c_2| = |c_1 - c_2| \tag{3.11}$$

In this case, for any measurement outcome at A there are equal chances for either the same or the opposite outcome at B. If, in addition, $\eta = 1$, we will have maximal entanglement and zero correlations.

As seen from these examples, correlation and entanglement are generally quite different concepts. The former may extend into classical domain, while the latter is purely QM phenomenon (see, e.g., [20-22]). The former is basis-dependent, while the latter is invariant under any rotations in $\mathcal{H}$.

(IV) States (3.1, 2) cannot be explained in terms of hidden variables (Bell's theorem [5-13]).

For this article, the most essential feature of entangled states is that measurement gives an *instant result* for both ends of (AB) even if performed at only one end. This means an immediate change of the whole state of the system no matter how widely extended; and it does so in a very subtle way – by changing *probabilities* $c_\alpha$ to 1 or to 0. It does not involve any superluminal signaling (SS) [8-11, 21, 22, 29-35]. The measurement outcomes at *A* and *B*, even though intimately linked, are not in the cause-and-effect relationship (we use the italics *A*, *B* to denote *events* in history of the respective objects A, B). The whole phenomenon including the instant change at both ends under the measurement at one end *has no classical explanation*.

Now we will try to show the consistency of IR with SR for a system of entangled particles. The argument will be based on two fundamental facts.

1) An A-measurement in system (3.1) or (3.2) *immediately* affects B [10-13]. The word *immediately* refers, as in IR of one-particle state, to *any* RF. This has been confirmed in numerous experiments. Simultaneous actualization of (+) or $(-)$ correlated outcomes works without any signaling between A and B simply because the memory of their common origin remains imprinted in the whole system (AB) no matter how widely it spreads.

2) Relativity of simultaneity does not preclude disentanglement from being instant in all RF since the simultaneous change of the corresponding *local* states is generally *not identical* to the pair of *physical events* (*actual measurements!*).

As an example, consider two spin-entangled electrons A and B with the zero net spin, receding from a common source C which is stationary in a frame $\mathrm{K_C}$. Let observer Alice move with A (frame $\mathrm{K_A}$) while Bob follows B (frame $\mathrm{K_B}$). Suppose both perform spin measurement (events *A* and *B*) at one moment by clocks of $\mathrm{K_C}$. Then in $\mathrm{K_A}$, event *A* happens earlier than *B*, while their succession is opposite in $\mathrm{K_B}$. And yet the disentanglement is instantaneous in either frame! Each observer *immediately* learns about the partner's result from the outcome of her/his own measurement. If Alice got the outcome $\left|\uparrow\right\rangle_A$, she immediately knows Bob's future result $\left|\downarrow\right\rangle_B$ simply because by the initial condition, the net spin of the system must remain zero. This means that B must disentangle to definite state $\left|\downarrow\right\rangle_B$ simultaneously with her measurement. The same holds for measurement's outcome $\left|\downarrow\right\rangle_A$: it immediately informs Alice about Bob's impending record $\left|\uparrow\right\rangle_B$. In either case, Bob's *actual* measurement merely confirms her prediction. The corresponding prediction by Bob is equally true in his RF. And there is no conflict between the two observers' statements. In $\mathrm{K_A}$, electron B disentangles (virtual "event" $B'$) at the moment of A-measurement. At this moment, there are no immediately observable occurrences at $B'$ since it is just change of the amplitudes in (3.1, 2)

$$(c_1, c_2) \to (1,0) \qquad (3.12)$$

or

$$(c_1, c_2) \to (0,1) \ , \qquad (3.13)$$

which will only later be confirmed by a B-measurement (actual *event B*). And the reciprocal argument will hold for $\mathrm{K_B}$. The important requirement in the whole analysis is drawing a sharp distinction between changes (3.12, 13) *of the amplitudes* with their respective *probabilities* (virtual "events" $A'$ and $B'$), and *actual measurements* (real events *A* and *B*). The first are simultaneous in any RF; the second may be simultaneous in $\mathrm{K_C}$ but time-separated in other RF.

There are two crucial elements in the physical structure of QM that require virtual events $A'$ and $B'$ to be simultaneous in *any* RF. The first one is unitarity. It precludes situations when, for example, instead of IR (3.12, 13) we will have in $\mathrm{K_A}$ first $(c_1, c_2) \to (1, c_2)$ at the moment of A-measurement and only later, at B-measurement, the final collapse to $(1, c_2) \to (1, 0)$. In such case of two-staged reconfiguration, unitarity would be violated during the whole time interval between the stages.

The second element is the conservation laws: they would be violated if, e.g., the zero net spin of bipartite AB becomes, at least temporarily, non-zero after the measurement on one of the particles, be it A or B.

The arguments are essentially the same as in the wave packet collapse. Technically, $B'$ is a point in space-time, and line $AB'$ is a simultaneity line in $\mathrm{K_A}$ (Fig. 1a). In Bob's frame, simultaneity line is $A'B$ (Fig. 1b). It is only in $\mathrm{K_C}$ that both measurements (events $\boldsymbol{A}$, $\boldsymbol{B}$, denoted hereafter in bold) are simultaneous (Fig. 1c).

We have here three different cuts through spacetime, each being simultaneity cut in its respective frame: $\boldsymbol{AB'}$ in $\mathrm{K_A}$, $\boldsymbol{A'B}$ in $\mathrm{K_B}$, and $\boldsymbol{AB}$ in $\mathrm{K_C}$. Each cut is the face of disentanglement in the corresponding frame. For *actual events*, the respective cut would be instantaneous in one RF and time-extended in others. But for *probabilities*, the world is insensitive to choice of the cut. This makes disentanglement an instant process in *any* RF without contradicting SR.

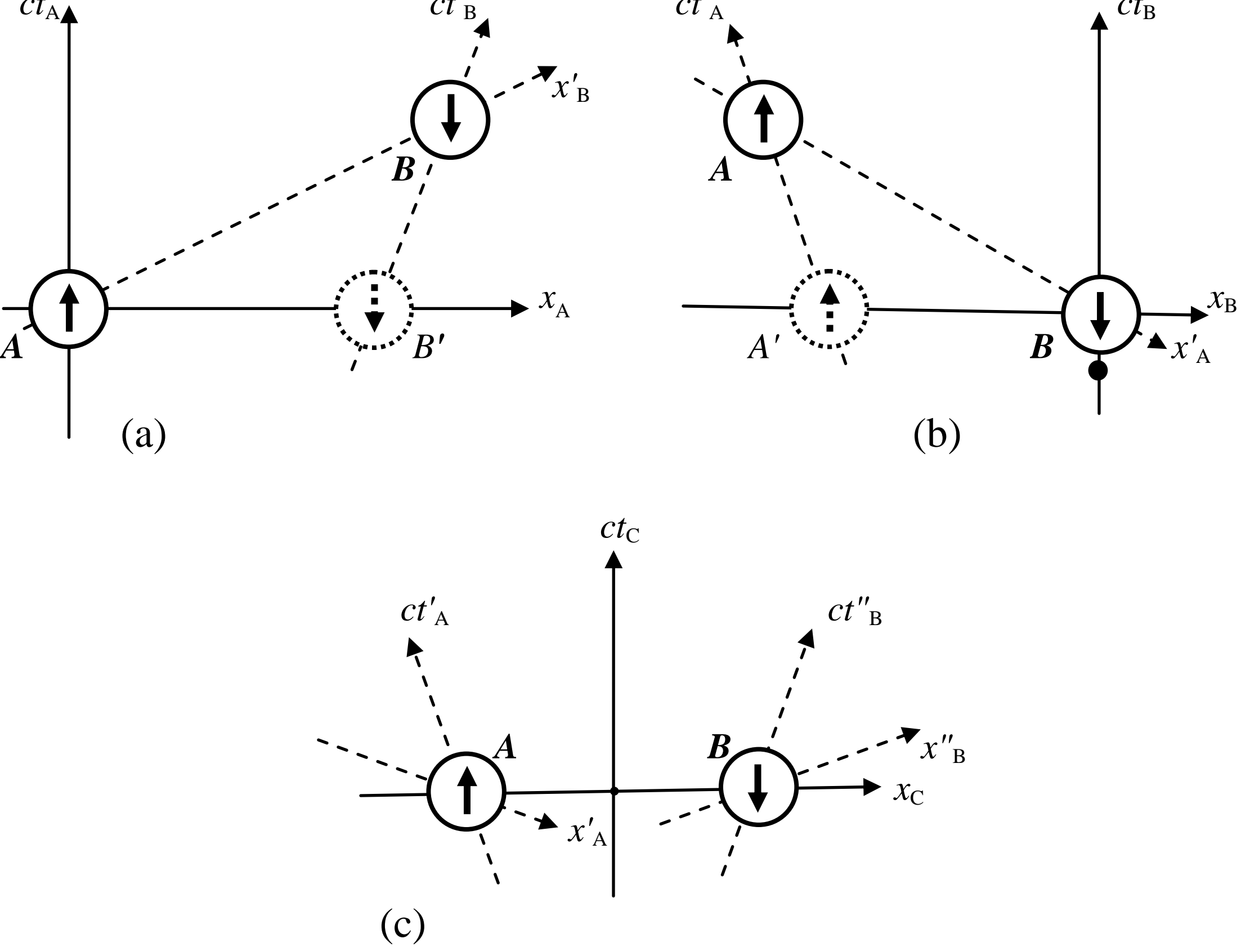

**Fig. 1**

Space-time diagrams (not to scale) for measurement of an entangled system in different RF.

(a) Frame $K_A$ with the origin at ***A***, simultaneity line ***A*** *B′*, and the time axis ***A***$ct_A$. The B gets disentangled (virtual event *B′***,** dotted arrow in the circle) simultaneously with ***A***, but the actual ***B***-measurement (event ***B***) is performed later than *B*'.

(b) Frame $K_B$ with the origin at ***B***, simultaneity line *A′* ***B***, and the time axis ***B***$ct_B$. The A gets disentangled (virtual event *A′* ) simultaneously with ***B*** but the actual A-measurement (event ***A***) is performed later than *A*'.

(c) Frame $K_C$ attached to the center of mass of (A, B). Events ***A*** and ***B*** here are simultaneous.

Geometrically, point $B'$ is an oblique projection of ***B*** onto the simultaneity line of $K_A$, and $A'$ is an oblique projection of ***A*** onto simultaneity line in $K_B$. These projections represent the fact that *disentanglement* caused by measurement at one end is an instant process in any RF, since the change at non-measured end is a *virtual event* which is by itself not directly observable until the actual measurement.

The whole argument holds even if the interval (***AB***) is time-like. Then one of the measurements, say, ***A***, precedes **B** in *any* RF, but disentanglement is still an instant process universally simultaneous with ***A*** in all frames. This holds even if the ***B***-measurement will be performed in the infinite future, which would practically mean measurement at only one end.

Thus, IR is a common feature of measurement of any spatially-extended QM object, be it a single particle or many-particle system. There may be some variations when we measure different characteristics of the two systems. In particle's position measurement, all possible outcomes form an infinite continuous set; in the spin measurement of a bipartite, we generally have a discrete set of 4 outcomes like in Eq. (3.6). In a certain basis, this set reduces to only two possible outcomes like in (3.1), (3.2). There is also subtle distinction in connections between actual events and their probabilities. In position measurement, out of two extremes (0 or 1) of local probability $d\mathcal{P}(\mathbf{r},t) = |\Psi(\mathbf{r},t)|^2 d\mathbf{r}$ within a volume element $d\mathbf{r}$ , only jumping to 1 is an immediately detectable event. In an entangled pair, we usually measure an observable like $s_z$ instead of position, and jump of probability $|c_\alpha|^2$ in (3.12, 13) to *any* of its extremes (0 or 1) is an event, but only at the measured end. At the opposite end, the concurrent jump generally falls short of being a recorded event until the actual measurement at that end. But at the same time the definite outcome is already known and is destined to be confirmed in a future measurement. Therefore it seems appropriate to use the term "*virtual event*" for points $A', B'$ in Fig.1, as was done in the text.

Very informative discussions of closely related issues without using abstract algebra, density matrix formalism, etc., can be found in [36, 37].

## 4. Conclusions

*The real world with quantum-mechanical indeterminacy can be relativistic only by being intrinsically probabilistic.* Therefore this article's title could as well be "*Quantum Mechanical Probabilities as the Allies of Relativity*". And with all that, the QM indeterminacy is deterministic because all the relevant *probabilities* are determined by QM to a highest possible accuracy, while the familiar determinism for actual *observables* is merely the classical limit of QM.

## Acknowledgements

I am grateful to Art Hobson for valuable comments and to Anwar Shiekh for inspiring discussions.